\documentclass[useAMS,letters]{mn2e}
\usepackage{graphicx}
\usepackage{epstopdf}
\def\gsim{\mathrel{\raise.5ex\hbox{$>$}\mkern-14mu
             \lower0.6ex\hbox{$\sim$}}}
\def\lsim{\mathrel{\raise.3ex\hbox{$<$}\mkern-14mu
             \lower0.6ex\hbox{$\sim$}}}

\newcommand{\aap}{    {\it Astron. Astrophys.}}

\newcommand{\apj}{    {\it Astrophys. J.}}

\newcommand{\mnras}{  {\it Mon. Not. Roy. Astron. Soc.}}

\newcommand{\nata}{    {\it Nature Astron.}}

\newcommand{\ssr}{    {\it Space Sci. Rev.}}

\title[MAW]{Magnetically Advected Winds}
\author[Contopoulos]{I. Contopoulos$^{1,2}$, D. Kazanas$^{3}$, and K. Fukumura$^{4}$\\
$^{1}$Research Centre for Astronomy and Applied Mathematics,
Academy of Athens, Athens 11527, Greece\\
$^{2}$National Research Nuclear University, 31 Kashirskoe highway, Moscow 115409, Russia\\
$^{3}$NASA Goddard Space Flight Centre, Greenbelt, MD 20771, USA\\
$^{4}$Physics \& Astronomy Department, James Madison University, Harrisonburg, VA 22840, USA}

\begin{document}

\date{Accepted ... Received ...; in original form ...}

\pagerange{\pageref{firstpage}--\pageref{lastpage}} \pubyear{2015}

\maketitle

\label{firstpage}

\begin{abstract}
Observations of X-ray absorption lines in magnetically driven disc winds around black hole binaries and active galactic nuclei yield a universal radial density profile $\rho\propto r^{-1.2}$ in the wind. This is in disagreement with the standard Blandford \& Payne profile $\rho_{\rm BP}\propto r^{-1.5}$ expected when the magnetic field is neither advected nor diffusing through the accretion disc. In order to account for this discrepancy, we establish a new paradigm for magnetically driven astrophysical winds according to which the large scale ordered magnetic field that threads the disc is continuously generated by the Cosmic Battery around the inner edge of the disc and continuously diffuses outward.
%The rate of magnetic flux diffusion is characterized by the magnetic diffusivity in the midplane of the disc.
We obtain self-similar solutions of such magnetically advected winds (MAW) and discuss their observational ramifications.
\end{abstract}


\section{Where does the disc magnetic field come from?}

The standard paradigm for the origin, acceleration and collimation of astrophysical jets and winds in black hole binaries
and active galactic nuclei involves the establishment of a large scale ordered magnetic field that threads the accretion disc around the central black hole. The magnetic field plays an important role not only in the acceleration and collimation of the wind, but also in the dynamics and evolution of the accretion disc itself. Its origin remains, however, uncertain. 

In studies of steady-state magnetically driven disc winds (Blandford \& Payne~1982 and subsequent works like Contopoulos \& Lovelace, etc.) it is usually implied that no large scale magnetic flux $\Psi$ is either advected inward or outward, namely that $\dot{\Psi}=0$. Equivalently,
\begin{equation}
E_\phi\equiv \frac{1}{c}(v_\theta B_r-v_r B_\theta)=0\ ,
\end{equation}
where, ${\bf v}$, ${\bf B}$, and ${\bf E}$ are the flow velocity, the magnetic and the electric field in the disc wind respectively, and $c$ is the speed of light (Chandrasekhar~1956). This scenario has been investigated in great depth by the group led by Jonathan Ferreira and Guy Pelletier who in the last twenty years studied the very origin of the magnetically driven wind in the diffusive part of the disc (Ferreira \& Pelletier~1993a, 1993b; Ferreira et al.~2006). That investigation was limited to self-similar disc-wind configurations. Nevertheless, the most important conclusion of their work has been that under the assumptions of steady-state and no flux advection, the radial density distribution in the wind cannot differ significantly from the canonical distribution assumed in the seminal paper of Blandford \& Payne~(1982), namely that
\begin{equation}
\rho_{\rm BP}\propto r^{-1.5}
\label{n}
\end{equation}
(in the nomenclature of Ferreira et al., $\rho\propto r^{p-1.5}$, and $p$, their `local ejection efficiency parameter' cannot differ much from zero).

This theoretical result is very important, but detailed observations of X-ray absorption lines (XRAL) in black hole binaries and active galactic nuclei imply a {\em different} density scaling in the wind, namely
\begin{equation}
\rho\propto r^{-1.2}
\label{nnew}
\end{equation}
(Fukumura et al.~2010a, 2010b, 2014, 2015), which in fact seems to be universal (Fukumura et al.~2017). It is interesting that this most recent observational result led Chakravorty et al.~(2017) to reconsider one of the original assumptions of the Ferreira et al. model, and to conclude that a standard Blandford \& Payne wind can only be launched from a `warm' (not `cold') disc. 

We believe that the disagreement between eqs.~(\ref{n}) and (\ref{nnew}) teaches us something more fundamental. The scenario where the magnetic field is held in place by accretion against its outward diffusion through the disc is problematic because at large distances the accretion disc is cold and highly non-ionized, thus also highly magnetically diffusive. This is often expressed by the magnetic Prandtl number (the ratio of viscosity $\nu$ to magnetic diffusivity $\eta$ in the midplane of the disc) being less than unity at large distances, namely ${\cal P}_{\rm m}\equiv \nu/\eta<1$ (e.g. Balbus \& Henri 2008). The disc is highly ionized, and its magnetic Prandtl number is greater than unity close to the central compact object. Nevertheless, it is unlikely that the magnetic field responsible for driving the disc wind is brought in from large distances (e.g. from a nearby star; Lubow et al.~1994). It is also unlikely that it is held in place.

We here would like to investigate a magnetically driven disc wind configuration in which the magnetic field is continuously generated around the inner part of the disc and from there diffuses outward. Such a scenario arises naturally in the theory of the so-called Cosmic Battery (Contopoulos \& Kazanas~1998).

\section{Magnetic flux generation and outward flux diffusion in the disc}

The Cosmic Battery is based on the simple fact that the electrons in the innermost optically thin plasma in orbit around the central compact object are deccelerated by the aberrated ambient radiation pressure, whereas the plasma ions are not. This effect generates an azimuthal electric current which in its turn generates poloidal magnetic field loops somewhere around the inner edge of the disc. These magnetic field loops open up by the differential disc rotation, and their inner parts are advected towards the centre, whereas their outer parts diffuse through the outer partially ionized diffusive part of the accretion disc (see Contopoulos et al.~2015, 2017 for numerical simulations of this process in action). 

The evolution of the magnetic field in the inner and outer disc is dictated by the induction equation expressed in spherical coordinates $(r,\theta,\phi)$ in the midplane of the disc as
\begin{equation}
\frac{\partial B_\theta}{\partial t}=-\frac{1}{r}\frac{\partial}{\partial r}\left[rc E_{\rm CB}+rv_r B_\theta-\eta\left(\frac{\partial (rB_\theta)}{\partial r}-\frac{\partial B_r}{\partial \theta}\right)\right]
\label{induction}
\end{equation}
(due to symmetry, $B_r=B_\phi=0$ in the midplane of the disc, but $\partial B_r/\partial\theta\neq 0$). Here, $v_r$ is the inward accretion disc
 velocity, and $\eta$ is the magnetic diffusivity in the midplane of the disc. $E_{\rm CB}$ is the azimuthal electric field generated by the Cosmic Battery, namely 
\begin{equation}
E_{\rm CB}=-\frac{L\sigma_{\rm T}}{4\pi r_{\rm g}^2 ce}{\cal F}(r/r_{\rm g})\ .
\label{ECBapprox}
\end{equation}
Here, $ r_{\rm g}\equiv GM/c^2$, $M$ is the mass of the central object, $L$ is the central luminosity, $m_{\rm p}$ is the proton mass, $e$ is the magnitude of the electron charge, and $\sigma_{\rm T}$ is the Thomson electron cross section. The expression in ${\cal F}(x)$ incorporates the effects of the abberation of radiation, $v_\phi/c$, the decrease of the radiation force as we approach the black hole horizon, and the high optical depth beyond some distance where the central radiation is heavily absorbed. ${\cal F}(x)$ can only be obtained through a detailed calculation of the radiation field in the vicinity of the central black hole in the optically thin part of the disc (Koutsantoniou \& Contopoulos~2014, 2017, and Contopoulos et al.~2017). One can also use simple analytical approximations for ${\cal F}(x)$ as in Contopoulos, Nathanail \& Katsanikas~(2015). Notice that under the convention that the direction of the disc angular velocity vector is along $\theta=0$, $E_{\rm CB}$ is everywhere negative, while $B_\theta$ is negative around the centre and positive at larger distances. 

In a steady rotational flow, the source term $E_{\rm CB}$ in eq.~(\ref{induction}) operates continuously and contributes to the growth of the magnetic field. Eventually, the two other terms become dominant. As we discussed above, near the central black hole, the dominant term is the second one in the r.h.s. of eq.~(\ref{induction}) corresponding to inward field advection. Further out, ${\cal P}_{\rm m}<1$ and thus the dominant term is the third one corresponding to outward field diffusion through the disc. We thus expect that around a distance $r_0$ on the order of a few times $r_{\rm g}$, the Cosmic Battery source term dictates the rate of total magnetic flux generation interior to that distance, namely
\begin{eqnarray}
\dot{\Psi}_{\rm CB} & \equiv & -2\pi r_0 cE_{\rm CB}\approx\frac{L\sigma_{\rm T}v_\phi(r_0)}{2 r_0 e}\nonumber\\
& \approx & 
2\pi\ \left(\frac{L}{L_{\rm Edd}}\right)\left(\frac{r_0}{r_{\rm g}}\right)^{-\frac{3}{2}}\frac{m_{\rm p}c^3}{e}\ ,
\label{Psidot}
\end{eqnarray}
where, $L_{\rm Edd}\equiv 4\pi GM m_{\rm p} c/\sigma_{\rm T}$ is the Eddington luminosity. According to the Cosmic Battery, the innermost polarity of this magnetic flux is held around the centre by the accreting highly ionized hot part of the disc, while the outer reverse polarity diffuses fast outwards. 

There has been a lot of work done around advection and diffusion of large scale magnetic fields in accretion discs. It appears clearly that the global outcome depends not only on the disc midplane conditions but also on the vertical disc stratification (e.g. Guilet \& Ogilvie 2013, Lovelace et al.~2009). 
In particular, $\partial B_r/\partial\theta\approx B_{rs}r/h$, where $B_{rs}$ is the value of $B_r$ on the surface of the disc which can only be obtained via a global calculation, and $h$ is the disc scale height. In cold discs where $h\ll r$, the last term in eq.~(\ref{induction}) is probably dominant. We thus estimate that 
\begin{equation}
\dot{\Psi}\sim 2\pi\eta B(r/h)\ ,
\label{diff}
\end{equation}
which is equivalent to a radial outward diffusion velocity through the midplane of the disc
 equal to
\begin{equation}
v_{\rm diff}\sim \frac{\eta}{h}\ .
\label{vdiff}
\end{equation}
The condition that ${\cal P}_{\rm m}\ll 1$ in the outer disc yields $|v_r|/v_{\rm diff}\ll 1 $. Notice that both $v_r$ and $v_{\rm diff}$ decrease to zero as $r$ increases to infinity.

The inward flux advection and accumulation cannot operate for ever, and at some finite interval of time the accumulated magnetic field will disrupt the innermost accretion process. During that time, magnetic field of the reverse polarity of that accumulated around the centre will `steadily' diffuse outward\footnote{Hereafter, we will use the term `steady' with quotation marks to denote a process that operates steadily for a finite interval of time.}. Therefore, we may assume that, in `steady' state, the rate of outward field diffusion will be constant over a range of distances. If we knew the radial distribution of $\eta$ in the disc, we could in principle integrate eq.~(\ref{diff}) above to obtain the radial distribution of the large scale magnetic field that threads the accretion disc, namely $B(r)\sim \dot{\Psi}h/2\pi \eta(r)r$ in `steady' state. However, the origin of the magnetic disc diffusion in the disc is not well determined (see below). We remind the reader that the corresponding radial distribution in the Blandford \& Payne wind model is
\begin{equation}
B_{\rm BP}\propto r^{-\frac{5}{4}}\ .
\label{BP}
\end{equation}

It is important to emphasize once again that, according to the scenario of the Cosmic Battery, the magnetic field that threads the disc represents the return (reverse) polarity of the magnetic field accumulated around the central black hole. As we argued in Contopoulos \& Kazanas~(1998), the maximum possible amount of magnetic flux that can be accumulated around the central black hole may be estimated as
\begin{equation}
\Psi_{\rm max}\sim r_{\rm ISCO}^2 B_{\rm eq}\sim 10^{20}\ \dot{m}^{0.5}(M/M_\odot)^{1.5}\mbox{cm}^2\ \mbox{G}\ \ .
\label{Psitotal}
\end{equation}
Here, $B_{\rm eq}\sim 10^8\ \dot{m}^{0.5}(M/M_\odot)^{-0.5}$~G is a rough estimate of the so-called equipartition magnetic field surrounding the black hole; $\dot{m}$ is the disc accretion rate in units of the Eddington accretion rate $\dot{M}_{\rm Edd}\equiv L_{\rm Edd}/c^2$. We also took $r_{\rm ISCO}\sim 6 r_{\rm g}$. Therefore, $\Psi_{\rm max}$ represents also the maximum amount of magnetic flux contained in the disc wind. We expect that some day we will be in a position to test this prediction of the Cosmic Battery observationally. 

\section{Magnetically driven winds with outward flux advection}

We will now solve the full set of steady-state ideal MHD equations with `steady' outward flux advection and thus show how the radial scaling dictated by eq.~(\ref{nnew}) may be implemented in nature. We will name these solutions magnetically advected winds (hereafter MAW). These differ from the standard steady-state ideal magnetically driven winds in which the flow and magnetic flux surfaces coincide and the equations that describe the dynamics of the flow and the magnetic field are combined into one central equation, the well known Grad-Shafranov equation. Now `steady' state implies a `steady' outward flux advection, namely $\dot{\Psi}=\mbox{const.}\neq 0$, or equivalently
\begin{equation}
E_\phi= \frac{1}{c}(v_\theta B_r-v_r B_\theta)= -\frac{\dot{\Psi}}{2\pi rc\sin\theta}\neq 0\ .
\label{Ephi}
\end{equation}
Contopoulos~(1996) showed that the full set of MHD equations is amenable to a radially self-similar solution where the main physical quantities (density, magnetic field and flow velocity) scale respectively as
\begin{eqnarray}
\rho(r,\theta) & = & \rho(\theta)(r/r_0)^{-w}\label{1}\\
{\bf B}(r,\theta) & = & {\bf B}(\theta)(r/r_0)^{-(2+w)/4}\label{2}\\
{\bf v}(r,\theta) & = & {\bf v}(\theta)(r/r_0)^{-(2-w)/4}\label{3}\ .
\end{eqnarray}
$r_0$ is some characteristic distance. Notice that, for simplicity of notation, we will henceforth be using the same symbols for the full physical quantities and their $\theta$ functional dependences.
The above radial scalings arise from dimensional arguments: $\rho$ must scale as $B^2/v^2$, $vB$ must scale as $E$, and  $E$ scales as $1/r$ (eq.~\ref{Ephi}). Thus, if one assumes a particular radial scaling for $\rho$ (e.g. the one in eq.~\ref{nnew} implied from observations of XRAL), the radial scalings of velocities and magnetic fields are fixed. In particular, if the scaling of velocities is Keplerian, i.e. if $v\propto r^{-0.5}$, then `steady' state flux advection requires that $\rho\propto r^0$. In other words, the density profile $\rho\propto r^{-1.2}$ implied by XRAL observations requires {\it deviation} from Keplerian velocity scaling and Blandford \& Payne magnetic field scaling, namely $v\propto r^{-0.2}$ and $B\propto r^{-0.8}$. Notice that this is not the case in standard self-similar MHD winds without flux advection in which the radial self-similar scalings of velocities and magnetic fields are independent (e.g. Blandfrod \& Payne~1982, Contopoulos \& Lovelace~1994, Ferreira \& Pelletier~1993, etc.). Obviously, in order to obtain general self-similar MAW solutions, we must ignore the gravity of the central object. Such an approximation is justified at large distances where the Keplerian velocity decreases faster than the MAW velocity.

Given the above scalings, we are now able to solve the full set of steady-state axisymmetric ideal MHD equations that describe ideal magnetically driven cold (zero pressure) winds with outward flux advection, namely
\begin{eqnarray}
&&\nabla\cdot(\rho{\bf v}) = 0\label{Phi}\\
&&\nabla\cdot({\bf B}) = 0\label{Psi}\\
&&\nabla\times({\bf v}\times{\bf B}) = 0\label{E}\\
&&4\pi\rho ({\bf v}\cdot\nabla){\bf v} = (\nabla\times {\bf B})\times {\bf B}\ .\label{force}
\end{eqnarray}
The equations of mass and magnetic flux conservation (eqs.~\ref{Phi} \& \ref{Psi} respectively) allow us to introduce flux functions $\Phi\equiv\Phi(\theta)(r/r_0)^{(6-3w)/4}$ and $\Psi\equiv\Psi(\theta)(r/r_0)^{(6-w)/4}$ in terms of which we express the poloidal flow and magnetic field components as
$\rho{\bf v}_{\rm p} \equiv \nabla\times (\Phi\hat{\phi}/2\pi r\sin\theta)$ and ${\bf B}_{\rm p} \equiv \nabla\times  (\Psi\hat{\phi}/2\pi r\sin\theta)$ respectively. It can easily be shown that
\begin{eqnarray}
v_r & = & \Phi_{,\theta}/2\pi\rho r_0^2\sin\theta
\label{vr1}\\
v_\theta & = & -(6-3w)\Phi/8\pi\rho r_0^2\sin\theta\\
B_r & = & \Psi_{,\theta}/2\pi r_0^2 \sin\theta\\
B_\theta & = & -(6-w)\Psi/8\pi r_0^2 \sin\theta\ .
\end{eqnarray}
Here, $(\ldots)_{,\theta}$ denotes ordinary differentiation with respect to $\theta$.
The $r$ and $\theta$ components of eq.~(\ref{E}) together with eq.~(\ref{Ephi}) yield
\begin{equation}
\rho(\theta) = [\Phi\Psi_{,\theta}(6-3w)-\Psi\Phi_{,\theta}(6-w)]/8\pi \dot{\Psi} r_0^3\sin\theta
\label{rho1}
\end{equation}
Combining all of the above into the $\phi$ component of eq.~(\ref{E}) and the 3 components of eq.~(\ref{force}) we obtain a system of 4 equations linear in $\Phi_{,\theta\theta}$, $\Psi_{,\theta\theta}$, $v_{\phi,\theta}$ and $B_{\phi,\theta}$ (see the Appendix). We integrate this system of equations and obtain the unknown functions $\Phi(\theta)$, $\Psi(\theta)$, $v_\phi(\theta)$ and $B_\phi(\theta)$, all the way from the surface of the equatorial disc ($\theta=\pi/2$) to the axis of symmetry ($\theta= 0$). Finally, physical quantities of the problem are obtained in terms of these functions combined with the scalings derived in eqs.~(\ref{1})-(\ref{3}).

A word of caution is in order here. The integration is not straightforward because the determinant of the matrix of coefficients of the above linear system of equations is equal to $2v_\theta^2(B_\theta^2-4\pi\rho v_\theta^2)(B^2-4\pi\rho v_\theta^2)/\dot{\Psi}r_0^3\sin\theta$ and vanishes at the positions (angles) where $v_\theta^2 = B_\theta^2 / (4\pi\rho)$ and $v_\theta^2 = B^2 / (4\pi\rho)$. Here, $B^2\equiv B_r^2+B_\theta^2+B_\phi^2$. The latter conditions define the critical surfaces of the flow, namely the Alfv\`{e}n and the so-called ``modified'' fast magnetosonic surface respectively. We suggest at this point that the interested reader consults the theory of modified critical surfaces in self-similar axisymmetric ideal MHD flows developed by Contopoulos~1996 and Tsinganos et al.~1996. It was shown there that the flow critical surfaces lie at the positions where the flow velocity perpendicular to the directions of symmetry becomes equal to the velocity of the characteristic waves perpendicular to the directions of symmetry. This is indeed confirmed in our present problem in which the direction perpendicular to the directions of symmetry (the $r$ and $\phi$ directions) is the $\theta$ direction. The boundary conditions on the surface of the disc must be chosen very carefully so that the solution crosses the critical surfaces without discontinuities. 

One particular MAW solution for $w=1.2$ is shown in Fig.~1. On the left, we plot poloidal flow and magnetic field lines as red and black solid lines respectively. These are simply surfaces of constant $\Phi(r,\theta)$ and $\Psi(r,\theta)$. Length units are arbitrary. As expected, due to the steady outward advection of magnetic flux, flow lines cross magnetic field lines from inside outward. The Alfv\`{e}n surface is denoted with a blue line. On the right, we plot the Alfv\`{e}n, modified fast magnetosonic, and the usual fast magnetosonic Mach numbers (blue, black, red lines respectively) defined as $\sqrt{4\pi\rho}v_\theta/|B_\theta|$, $\sqrt{4\pi\rho}v_\theta/B$, and $\sqrt{4\pi\rho}v_{\rm p}/B$ respectively as functions of wind latitude $90^o-\theta$. 
%We also plot the poloidal flow velocity (dashed line) along the flow line originating at $r_0=1$ in units of the characteristic azimuthal velocity at its base.
The equatorial boundary conditions on the surface of the disc are $\Psi/2\pi=1$, $\Psi_{,\theta}/2\pi=0.94$, $\Phi/2\pi=0.51$, $\Phi_{,\theta}/2\pi=0.57$, $v_\phi=1$, and $B_\phi=-1$. These correspond to ${\bf v}=(0.14,-0.08,1)$ in arbitrary units, ${\bf B}=(0.94,-1.2,-1)$ in arbitrary units, and $\rho=4$ in $B^2/v^2$ units. $\Psi_{,\theta}$ is carefully chosen so that the flow crosses the Alfv\`{e}n surface smoothly. If we specify arbitrary physical units $r_0=10r_{\rm g}=300$~km for a $10M_\odot$ central black hole, $v_\phi(r_0)=0.1\ c$, $B(r_0)=10^6$~G, the above correspond to $\rho(r_0)=4\times 10^{-7}$~g\ cm$^{-3}$, $\dot{M}_{\rm wind}=10^{-9}(r/r_0)^{0.8}\ M_\odot/{\rm yr}$ (this yields $5\times 10^{-5}\ M_\odot/{\rm yr}$ over a radial distance of 1~AU in the disc), and $\dot{\Psi}\sim 5\times 10^{22}$~G\ cm$^2$/s, or equivalently $v_{\rm diff}\sim \dot{\Psi}/2\pi rB\sim 0.008\ c\ (r/r_0)^{-0.2}$.
% (eqs.~\ref{diff} \& \ref{vdiff}).

In this particular solution, the wind is accelerated to super-Alfv\`{e}nic and super -fast magnetosonic velocities, but remains everywhere sub-modified fast magnetosonic. The solution is thus not unique. Notice that this is also the case for most magnetically driven wind solutions in the literature (e.g. Blandford \& Payne~1982, Contopoulos \& Lovelace~1994, etc.). We plan to investigate the parameter space of possible solutions in a future publication.

\begin{figure}
\begin{minipage}[b]{0.45\linewidth}
\centering
\includegraphics[width=1.4\columnwidth]{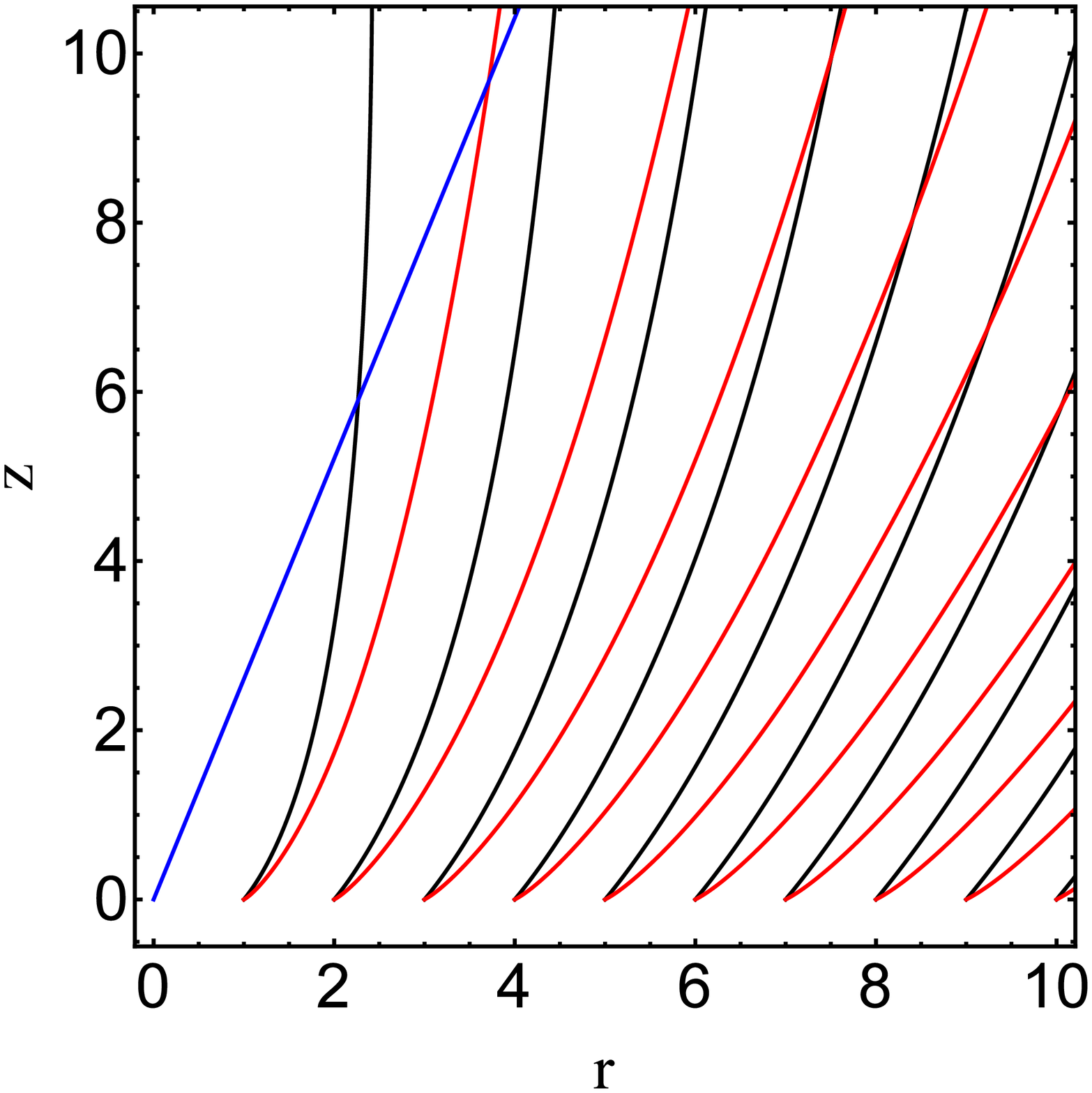}
%\includegraphics[trim=0cm 4cm 0cm 4cm,
%clip=true, width=4.5cm, angle=0]{Figure1.eps}
%\caption{blah blah} \label{fig:blah1}
\end{minipage}
\hspace{1.5cm}
\begin{minipage}[b]{0.35\linewidth}
\centering
\includegraphics[width=0.905\columnwidth]{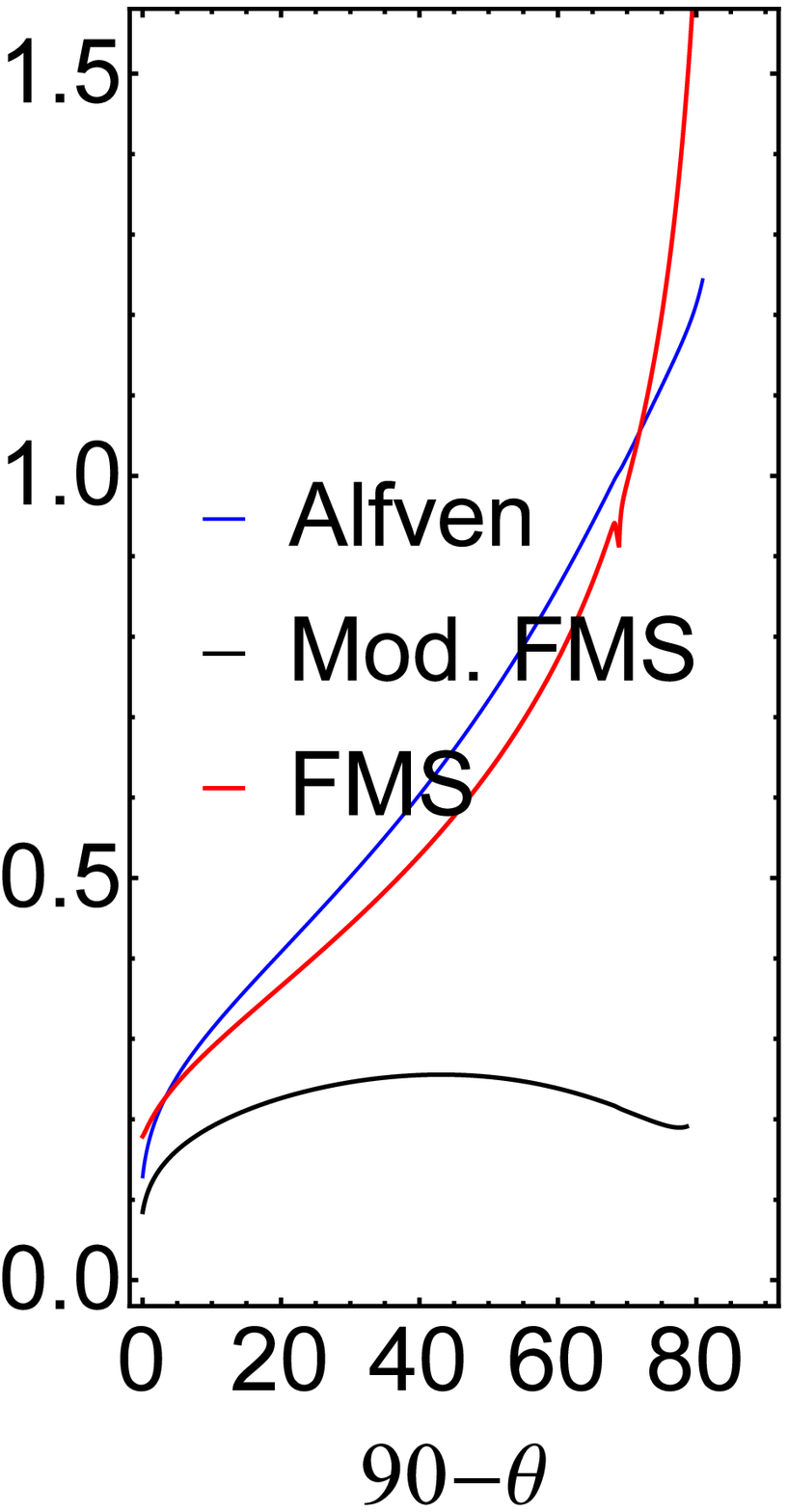}
%\includegraphics[trim=0cm 4cm 0cm 4cm,
%clip=true, width=4.5cm, angle=0]{Figure2.eps}
%\caption{blah blah} \label{fig:blah2}
%  \includegraphics[width=4cm, angle=90]{L11.eps}
\end{minipage}
  \caption{MAW solution for $w=1.2$. Left: Wind structure in arbitrary length units. Red/black lines: poloidal flow/magnetic field lines respectively. Blue line: Alfv\`{e}n surface. Right: Mach numbers as functions of wind latitude. Blue/black/red lines: Alfv\`{e}n/modified fast magnetosonic/fast magnetosonic Mach numbers respectively.}
 \label{fig}
\end{figure}

%\begin{figure}
%\centering
%\includegraphics[width=0.6\columnwidth]{Figure1.eps}
 %\caption{MAW solution for $w=1.2$. Thin solid lines: poloidal flow lines. Dashed lines: poloidal magnetic field lines. Thick solid line: Alfv\`{e}n surface. Length units are arbitrary}
%\label{figure1}
%\end{figure}

%\begin{figure}
%\centering
%\includegraphics[width=0.4\columnwidth]{Figure2.eps}
%\caption{Flow velocities and Mach numbers along the flow line originating at $r_0=1$ as functions of wind latitude. Solid line: Alfv\`{e}n Mach number. Dotted line: fast magnetosonic Mach number (see text for details). Dashed line: poloidal flow velocity in units of the characteristic azimuthal velocity at the base of the flow on the equatorial disc. Thick vertical solid line: position of the Alfv\`{e}n critical surface.}
%\label{figure2}
%\end{figure}

Our solution refers only to the ideal MHD wind, and does not consider the physical conditions in the disc, namely how the accretion flow manages to gradually divert some part of it into a MAW outflow. For $w=1.2$, the outward magnetic field diffusion velocity in the disc must scale radially as $v_{\rm diff}\propto r^{-0.2}$, and according to eq.~(\ref{vdiff}), $\eta\propto r^{0.8}$ in the disc (assuming $h/r\sim$~const). Notice that the standard expression for the Spitzer magnetic diffusivity in the midplane of the outer non-ionized part of the disc, yields a {\it different} radial scaling, namely $\eta\propto r^{9/8}$ (Spitzer~1962, Balbus \& Henri~2008). 
%This implies that, if we want to establish a radially self-similar %MAW solution above the disc, the magnetic diffusivity in the %disc {\em must differ from} the standard Spitzer magnetic %diffusivity.
One possible alternative is anomalous turbulent diffusivity (e.g. Lubow et al.~1994). We will return to this problem in the future and address the disc-wind connection, as did Ferreira et al. for the standard Blandford \& Payne-type MHD winds.

\section{Summary and Discussion}

In this work, we obtained a new type of magnetically driven disc wind solutions with outward flux advection (MAW). The reader should not be confused by the fact that the flow velocity ${\bf v}$ is not parallel to ${\bf B}$. The solutions are ideal and the wind plasma is `frozen in' the magnetic field, but magnetic field lines do not `stand still': not only do they rotate in the azimuthal direction as in all previous standard axisymmetric ideal MHD solutions, but they are also advected in the poloidal direction in such a way so that the magnetic field distribution remains unchanged with time (except for the region around the axis where magnetic field loops are generated). We plan to perform a more detailed investigation of our solutions in a future publication. 

As we argued earlier, MAW configurations cannot remain in steady-state for ever. Magnetic flux will be accumulated around the centre up to the point of equipartition, beyond which the magnetic field will severely alter the process of accretion. At that point, the accumulated field configuration will become unstable, and the whole process of inward flux advection and outward flux diffusion through the disc will start all over again. We thus expect that the resulting disc winds will not be established for ever, but will also follow a certain cycle of destruction and re-generation. The timescale for this process is of the same order as the timescale for growth of the central magnetic field to equipartition by the Cosmic Battery, namely
\begin{equation}
\tau_{\rm CB} \sim 10^2\ \dot{m}^{-0.5}
(M/M_\odot)^{1.5}\mbox{s}
\label{taueq}
\end{equation}
(eq.~12 of Contopoulos \& Kazanas~1998; see also Kylafis et al.~2012). This lifetime is on the order of weeks for solar mass black holes, and a billion years for supermassive black holes. We would like to end this work with a summary of our main conclusions. 
\begin{enumerate}
\item MAW solutions differ significantly from the standard Blandford \& Payne-type outflows. In particular, the magnetic field radial profile differs significantly from the standard Blandford \& Payne profile, and the velocity radial profile differs significantly from Keplerian. In fact, the analysis of 26 Seyfert galaxy warm absorbers by Laha et al.~(2014) yields $v\propto \xi^{0.12}$. Here, $\xi$, the wind ionization parameter, is proportional to $1/\rho r^2$, which further yields $v\propto r^{-0.1}$. This result is more consistent with our present MAW scaling $v\propto r^{-0.2}$ than the standard Keplerian one. However, the $v$ vs. $\xi$ correlation seems different from source to source, and $v\propto r^{-0.5}$ is also a good approximation in some cases (see e.g. Detmers et al.~2011 and Gupta et al.~2015 for Mrk 509). Still, the fact that the wind velocity radial profile often appears shallower than Keplerian makes the new solutions very interesting. 
\item Accretion discs with magnetic Prandtl number less than unity can only support large scale magnetic fields if the latter are generated by some mechanism operating around the central compact object and from there diffuse outward to large distances.
\item  The Cosmic Battery sets a limit on the duration of the return-flux diffusion through the disc, thus also on the duration of the magnetically driven disc wind. According to eq.~(\ref{taueq}), this timescale
is on the order of weeks in stellar mass black hole binaries, and on the order of a billion years in active galactic nuclei.
\item One corollary of the above argument is an estimate of the total amount of magnetic flux contained in the wind (eq.~\ref{Psitotal}). One could in principle test this observationally and compare it with corresponding estimates from standard jet theory.
\end{enumerate}

%\section*{Acknowledgements}
%We would like to thank the first referee who helped us improve an early version of the present work.

%\bibliographystyle{mnras}
%\bibliography{bibliography}

%\begin{thebibliography}{99}
%\end{thebibliography}

\section*{Appendix}

We introduce the scalings of eqs.~(\ref{1})-(\ref{3}) in the $\phi$ component of eq.~(\ref{E}) and the 3 components of eq.~(\ref{force}), and obtain a linear system of 4 equations
\begin{equation}
{\cal A}_{ij}{\cal X}_j={\cal B}_i
\end{equation}
for the unknowns ${\cal X}\equiv [\Phi_{,\theta\theta},\Psi_{,\theta\theta},(v_\phi)_{,\theta},(B_\phi)_{,\theta}]$. Indices take values $1,2,3,4$, and double indices imply summation from 1 to 4. The nonzero coefficients in the above system of equations are:
\[{\cal A}_{11} =-\frac{v_\theta B_\theta B_\phi}{\rho\dot{\Psi}r_0},\ 
{\cal A}_{12} = \frac{v_\theta^2 B_\phi}{\dot{\Psi} r_0},\ 
{\cal A}_{13} = -B_\theta,\ 
{\cal A}_{14} = v_\theta
\]
\[
{\cal A}_{21} = \frac{2v_\theta}{\sin\theta r_0^2}+\frac{4\pi v_r v_\theta B_\theta}{\dot{\Psi}r_0},\ 
{\cal A}_{22} = -\frac{B_\theta}{2\pi\sin\theta r_0^2}-\frac{4\pi\rho v_r v_\theta^2}{\dot{\Psi}r_0}
\]
\[
{\cal A}_{31} = \frac{4\pi v_\theta^2 B_\theta}{\dot{\Psi} r_0},\ 
{\cal A}_{32} = \frac{B_r}{2\pi\sin\theta r_0^2}-\frac{4\pi\rho v_\theta^3}{\dot{\Psi}r_0},\ 
{\cal A}_{34} = B_\phi
\]
\[
{\cal A}_{43} = 4\pi\rho v_\theta,\ 
{\cal A}_{44} = -B_\theta
\]
\begin{eqnarray}
{\cal B}_{1} & = & -\frac{6-3w}{4}v_r B_\phi+\frac{6-w}{4}v_\phi B_r \nonumber\\
&& +\frac{\pi w v_r v_\theta B_r B_\phi \sin\theta r_0}{\dot{\Psi}}+\frac{v_\phi B_\theta}{\tan\theta}\nonumber\\
{\cal B}_{2} & = & \frac{2-w}{4}B_\phi^2-\frac{2-w}{4}B_\theta^2-\frac{B_r B_\theta}{\tan\theta}\nonumber\\
&& +4\pi\rho\left(\frac{2-w}{4}v_r^2+v_\theta^2+v_\phi^2-\frac{\pi w v_r^2 v_\theta B_r\sin\theta r_0}{\dot{\Psi}}\right)\nonumber\\
{\cal B}_{3} & = & \frac{2-w}{4}B_r B_\theta +\frac{B_r^2-B_\phi^2}{\tan\theta}\nonumber\\
&& -4\pi\rho\left((w-1)v_r v_\theta+\frac{\pi w v_r v_\theta^2 B_r \sin\theta r_0}{\dot{\Psi}}\right)\nonumber\\
{\cal B}_{4} & = & \frac{B_\theta B_\phi}{\tan\theta}+\frac{2-w}{4}B_r B_\phi
+4\pi\rho\left(\frac{v_\theta v_\phi}{\tan\theta}+\frac{2+w}{4}v_r v_\phi\right)\nonumber
\end{eqnarray}
and ${\cal A}_{23}, {\cal A}_{24}, {\cal A}_{33}, {\cal A}_{41}, {\cal A}_{42}$ are equal to zero. ${\bf v}(\theta)$, ${\bf B}(\theta)$, and $\rho(\theta)$, are functions of $\theta$, $\Phi(\theta)$, $\Phi_{,\theta}(\theta)$, $\Psi(\theta)$, and $\Psi_{,\theta}(\theta)$, defined in eqs.~(\ref{vr1})-(\ref{rho1}). The parameter $\dot{\Psi}$ is defined in eq.~(\ref{Ephi}).

\end{document}